\documentclass[11pt,dvips]{article}
\usepackage{dcc}
\usepackage{epsfig,subfigure}

\begin{document}

\title{Using compression to identify acronyms in text}
\date{}
\author{{\em Stuart~Yeates\/}, {\em David~Bainbridge} and {\em
	Ian~H.~Witten\/} \\[1ex]
        Department of Computer Science \\
         University of Waikato \\ Hamilton, New Zealand \\[1ex]
        {\em \{s.yeates, d.bainbridge, i.witten\}@cs.waikato.ac.nz}}

\maketitle

\section{Introduction}

Text mining is about looking for patterns in natural language text, and may be
defined as the process of analyzing text to extract information from it for
particular purposes.  In previous work, we claimed that compression is a key
technology for text mining, and backed this up with a study that showed how
particular kinds of lexical tokens---names, dates, locations, {\em etc}.---can
be identified and located in running text, using compression models to provide
the leverage necessary to distinguish different token types (Witten {\em et
al.\/}, 1999).

Identifying acronyms in documents---which is certainly also about looking for
patterns in text---presents a rather different kind of problem.  Webster
defines an ``acronym'' as
\begin{quote}
a word formed from the first (or first few) letters of a series of words, as
{\em radar\/}, from {\em ra\/}dio {\em d\/}etecting {\em a\/}nd {\em
r\/}anging.
\end{quote}
Acronyms are often defined by preceding (or following) their first use with a
textual explanation---as in Webster's example above.  Finding all the acronyms,
along with their definitions, in a particular technical document is a problem
of text mining that has previously been tackled using {\em ad hoc\/}
heuristics.  The information desired is relational, comprising both acronyms
and their definitions, and this makes it rather different from the token-type
problem mentioned above.

It is useful to detect acronyms in an information retrieval or digital
library context for several reasons.  First, new tools could be built
from the data gathered across a document collection---tools such as
browsable acronym lists and search-by-acronym indexes.
Second, explicit recognition of acronyms could make existing tools
like search engines and keyphrase extraction schemes work more
effectively by performing in-line substitutions.  Third, reading
acronym-laden source documents could be enhanced by annotating text
with pop-up balloons that recall the definitions of acronyms, or menus
that help the user navigate to other documents that contain the
acronym.  Fourth, spell-checking is often applied as a text cleaning
operation after textual data capture using OCR, and acronym detection
improves the quality of this process.

It is not immediately obvious how compression can assist in locating relational
information such as acronyms and their definitions.  Language statistics of
acronyms will certainly differ from those of ordinary running text, because
acronyms have a higher density of capital letters, and a far higher density of
non-initial capital letters.  However, other words are also capitalized, and it
seems unlikely that acronyms will be recognized reliably on this basis.
Moreover, we are interested not just in the occurrence of acronyms, but their
definitions too, and these will certainly not be readily distinguished from
ordinary language by their letter statistics.


We have experimented with a simple way of coding potential acronyms with
respect to the initial letters of neighboring words, and using the compression
achieved to signal the occurrence of an acronym and its definition.  This seems
to form the basis for a reliable acronym finder.  The next section discusses
acronyms themselves, and distinguishes them from ordinary abbreviations.
Following that we examine three existing systems for acronym extraction from
text.  These rely heavily on heuristics that are provided {\em a priori\/} by
the system designer, based on the designer's experience and skill.  In
Section~\ref{sec:compression} we introduce a new algorithm for
compression-based acronym identification, and in Section~\ref{sec:results} we
compare this algorithm to existing schemes.

\section{Acronyms}

Abbreviations are contractions of words or phrases that are used in place of
their full versions, where their meaning is clear from the context.  Acronyms
are a type of abbreviation made up of the initial letter or letters of other
words.  Key differences between acronyms and other abbreviations include the
lack of symbols such as apostrophe and period in acronyms, a more standard
construction, and the use of capital letters.  {\em Can't\/} and {\em etc.\
}are abbreviations but not acronyms, in the first case both because of the
inclusion of other than initial letters and because of the inverted comma, and
in the second case because of the use of a period; moreover, both contain
lower-case letters.

Very commonly used and easily pronounced acronyms may enter the language as
bona fide words, in which case they treated like regular nouns and only
capitalized at the beginning of sentences.  {\em Laser\/} (for Light
Amplification by Stimulated Emission of Radiation) and {\em radar \/} (for
RAdio Detection And Ranging) are such words.

Acronyms are a relatively new linguistic feature in the English language, first
appearing in the 1940s and 1950s.  Early acronyms include {\em AmVets\/} for
American Veterans' Association (1947), {\em MASH\/} for Mobile Army Surgical
Hospital (1954) and {\em MASER\/} for Microwave Amplification by Stimulated
Emission of Radiation (1955).  Fertile sources of acronyms have been
organizational ({\em UNESCO\/}, {\em VP\/}, {\em CEO\/}), military ({\em
MASH\/}, {\em SIGINT\/}, {\em NATO\/}) and scientific jargon ({\em laser\/},
{\em ROM\/}, {\em GIS\/}).

Acronyms may be nested, for example {\em SIGINT\/} ({\em Signals
Intelligence\/}) is part of {\em JASA\/} ({\em Joint Airborne SIGINT
Architecture\/}) or even recursive as in {\em GNU\/} ({\em GNU's Not Unix\/}).
Recursive acronyms appear to be constructed self-consciously and are limited to
a few domains.  Although mutually recursive acronyms appear possible, they seem
unlikely.

Acronym lists are available from several sources,\footnote{e.g. {\em
www.geocities.com/\~{}mlshams/acronym/acr.htm} and {\em www.ucc.ie/acronyms}}
but these are static---they list acronyms current in some domain at the time of
compilation or officially endorsed by an organization.  While these may be of
use in specific contexts, they are unlikely to be useful for an arbitrary piece
of text at some point in the future.

Abbreviations such as acronyms are used in places where either readers
are familiar with the concepts and entities they stand for or their
meanings are clear from the context of the discussion.  Unlike other
abbreviations, acronyms are usually introduced in a standard format
when used for the first time in a text.  This form is: {\em ROM (Read
Only Memory)\/} or alternatively {\em Read Only Memory (ROM)\/}, the
latter being preferred when readers are unlikely to be familiar with
the concept.  Later instances of the acronym can be given simply as
{\em ROM}. 

Acronyms are not necessarily unique.  The Acronym Finder web site\footnote{{\em
www.mtnds.com/af/}} has 27 definitions for {\em CIA\/}, ranging from {\em
Central Intelligence Agency\/} and {\em Canadian Institute of Actuaries\/} to
{\em Central Idiocy Agency\/} and {\em Chemiluminescence Immunoassay}.  In
normal texts, non-uniqueness does not pose a problem: usually the meaning is
clear from the context of the document.  However, ambiguity is likely to be an
issue if acronyms are extracted from large, broad-based collections.
Extracting acronyms from clusters of related documents (grouped using document
clustering techniques) is one solution to this problem.

Acronyms are generally three or more characters in length, although
two-character acronyms exist (for example {\em AI} for {\em Artificial
Intelligence\/}).  Because of the small number of combinations, two-character
acronyms exhibit far greater scope for ambiguity (for instance {\em Artificial
Intelligence} versus {\em Artificial Insemination\/}).  Unless they refer to
very widely-known entities (e.g. {\em UN} for {\em United Nations\/} or {\em
EC} for {\em European Community\/}) they are generally restricted to a fairly
local scope---within a well-focused conceptual area, for example, or even
within a single document.

\section{Heuristic acronym extraction}

We sketch the operation of three automatic acronym detection programs that
exemplify different approaches to the problem.  One uses a longest common
subsequence method, based on the initial letters of neighboring words.  This
allows approximate matching (because not every letter in the acronym has to
originate in an initial letter of the definition) but makes it difficult to
incorporate additional letters, other than the first, from words in the
definition.  The remaining methods identify a source for every letter in the
acronym, which allows greater flexibility in determining the source of those
letters.

\subsection*{AFP: Acronym finding program}

AFP (Acronym Finding Program) was developed to improve post-processing of text
captured using OCR (Taghva and Gilbreth 1995).  Acronym candidates are defined
as upper-case words from three to ten characters in length.  The upper bound is
an arbitrary but reasonable assumption, while the lower bound is a compromise
between recall (as noted above, there are two-character acronyms) and precision
(approximate matching on anything less than three characters is error-prone).
Insisting that all acronyms appear in upper case
will cause many to be omitted---such as Webster's {\em radar\/} example above.

Acronym candidates are first tested against a list of ``reject words'' that
commonly appear in upper-case, such as {\em TABLE\/}, {\em FIGURE\/}, and roman
numerals.  For each candidate that passes this test, AFP constructs two text
windows, one containing the words that precede the candidate, the other
containing the words that follow.  In both cases the number of words in the
window is twice the number of characters in the acronym candidate. The sequence
of initial letters of these words is matched against the acronym itself, using
a standard longest common subsequence algorithm (Cormen {\em et al.\/}, 1993).

This can yield several candidates for the acronym definition. For example, in
\begin{quote}
management of the Office of Nuclear Waste Isolation (ONWI)
\end{quote}
the eight-word window contains three matches---two occurrences of ``of Nuclear
Waste Isolation'' and one of ``Office Nuclear Waste Isolation.''  In order to
decide which to return, the algorithm classifies words into {\em stop words\/},
{\em hyphenated words\/}, and {\em normal words\/}, and calculates a heuristic
score for each competing definition.  The calculation depends on the number of
normal words that must be skipped to make the acronym match (for example, using
the first {\em of\/} for ``of Nuclear Waste Isolation,'' the word {\em
Office\/} must be skipped to obtain a match), the number of stopwords used in
the acronym definition, the number of text words spanned by the acronym
definition, and the number of words that separate the acronym definition from
the acronym itself.  Effectively the shortest, closest candidate with the
lowest density of stop words is chosen.

\subsection*{TLA: Three-letter acronyms}


TLA (Three-Letter Acronyms) was developed to provide enhanced browsing
facilities in a digital library (Yeates 1999).  As with AFP, candidate acronyms
and their definitions are selected from a stream of words.  All non-alphabetic
characters are converted to spaces and any multiple spaces replaced with a
single space.

Candidate acronyms are determined by matching the initial letter of each word
in the context of a potential acronym against the appropriate letter in the
acronym.  If the first letter does not match, the word is skipped.  Otherwise,
the next letter of the same word is tested against the next letter of the
acronym, and if it matches the algorithm continues to move along the word.  A
maximum of six letters are used from each word, and a potential acronym must be
entirely upper-case.

In order to determine which candidate acronyms should be output, a machine
learning scheme is used.  Four attributes are calculated for each candidate:
\begin{itemize}
\setlength{\itemsep}{0cm}
\setlength{\parskip}{0cm}
\item
the length of the acronym in characters (generally between 2 and 6);
\item
the length of the acronym's definition in characters (generally between 10 and
40); 
\item
the length of the acronym's definition in words (generally between 2 and 6);
\item
the number of stop words in the acronym's definition.
\end{itemize}
These features clearly include redundancy---the fourth is the difference
between the third and the first.  The machine learning approach is to generate
a model using training data in which acronyms have already been marked by hand.
The model determines what attributes, and what combinations of attributes, are
the important ones for making the decision (Witten and Frank, 2000).  We used
the naive Bayes learning scheme from the Weka workbench, supplied with training
data as described below (Section~\ref{sec:data}).  The model produced by Naive
Bayes is then used to determine whether to accept a candidate acronym, on the
basis of the four features computed from it.

\subsection*{Perl Acronym Finder}

A third algorithm, designed for speed and high recall for a fairly small
collection of documents, was also available to us.
The documents in this collection were accurately represented and contained very
few errors. This algorithm, written in Perl, used the simple method of taking
the first letter from each word in a region around each candidate acronym
without accounting for any stopwords or errors, and matching the resulting
sequence against the candidate.

\subsection*{Discussion}

There are several important differences between the AFP and TLA algorithms.
\begin{itemize}
\item
AFP only considers the first letter of each word when searching for acronyms.
(Acronyms containing characters other than the first letter may be matched, if
the longest common subsequence algorithm ends up ignoring some characters of
the acronym.)  On the other hand, TLA considers the first six letters in each
word, which enables it to match acronyms such as {\em MUTEX} for {\em MUTual
EXclusion}.
\item
AFP parses words with embedded punctuation as single words, whereas TLA parses
them as separate words.  TLA's strategy allows matching of {\em U.S. Geographic
Service (USGS)\/} but may inhibit matching of other acronyms, although
no examples have been encountered so far.
\item
AFP tolerates errors, which sometimes matches acronyms that TLA misses.  For
example TLA misses {\em DBMS (DataBase Management System)\/} because the {\em
B} is embedded within {\em DataBase\/}, whereas AFP will obtain the
three-letter longest subsequence match {\em DMS} and infer the {\em B\/}.
\end{itemize}
Overall, TLA explicitly recognizes the need to tailor acronym extraction to the
particular problem domain, whereas AFP makes irrevocable decisions early on.
TLA seems to be more general than AFP in that a larger number of acronyms fall
within its scope.  On the other hand, the fact that AFP tolerates error works
in the other direction.

\section{Compression-based acronym extraction}
\label{sec:compression}

Can compression techniques be used as the basis for a text mining problem such
as acronym detection?  Our criterion is whether a candidate acronym could be
coded more efficiently using a special acronym model than it is using a regular
text compression scheme.  A word is declared to be an acronym if the ratio
between the number of bits required to code it using the acronym model is less
than a certain proportion of the number of bits required to code it in context
using a general-purpose compressor, and we experimented with different values
of the threshold.

We first pre-filter the data by identifying acronym candidates and determining
two windows for each, one containing preceding words and the other containing
following ones.  For our initial work we followed AFP's strategy of identifying
words in upper case as candidate acronyms (though we did not use a reject
list).  We chose a window containing the 16 preceding words, and a separate
window containing the 16 following ones---this covered all the acronym
definitions in our test data.

PPM is used as the reference text compression model (Bell {\em et al\/}.,
1990), with escape method D (Howard, 1993) and order 5 (order 6 yielded
slightly better performance on the training data, but only by a very small
margin).

\subsection{Coding the acronyms}

To code an acronym, its characters are represented with respect to the initial
letters of words in the window, and a string is produced that determines what
words, and what letters, those are.  Figure~\ref{fig:codings} shows examples,
with the acronym on the left, the text that defines it in the middle, and the
code on the right.  The first component of the code is whether the acronym
precedes (\verb.+.) or follows (\verb.-.) its definition, and the second is the
distance from the acronym to the first word of the definition.  The third is a
sequence of words in the text, each number giving an offset from the previous
word (for example, \verb.1. represents the next word of the text).  The fourth
gives the number of letters to be taken from each word (for example,
\verb.1. indicates just the first letter).

In the first example of Figure~\ref{fig:codings}, the acronym {\em BC\/} is
directly preceded by its definition.  From the acronym we go back (\verb.-.)
two words (\verb.2.) and use that word and the next (\verb.1.), taking one
character (namely the first) from each of the two words (\verb.<1,1>.).  In the
second example, the first character of the second word following the acronym
{\em OTC\/} is used, then the first character of the next word, and finally the
first character of the following word.  Case differences were ignored in these
experiments.

As the third example ({\em FG\/}) shows, several words are sometimes
interpolated between an acronym and its definition.  This also illustrates that
there is a fine line between acronyms and variable names, although from the
later context of the paper it is clear that {\em FG\/} is being used as an
acronym for the Fast Givens transform.  The largest distance between an acronym
and its definition that we encountered involves a word that is thirteen words
away from the acronym, shown as the fourth example {\em CHARME\/}---which is
why we use a much larger window than AFP.  This is also the first example that
takes more than one letter from a word: three letters come from {\em
hardware\/} and two from {\em methods}.  In the next example {\em CDAG\/},
three letters are taken from the second word---which is itself an acronym {\em
DAG\/}, nested in the acronym definition.  The {\em COMPCOM\/} example shows
four characters being taken from a word, which is rare.  More than four
characters did not occur in our data, and we imposed an upper limit of six on
the number of characters that could be taken from a word.  The largest acronym
that we know has eight letters ({\em WYSINWYG} for {\em what you see is not
what you get\/}), which can easily be coded since the window is 16 words long
and any number of these words can be used to make up the acronym.

In order to operate this coding scheme it is necessary to parse the input into
words.  Hyphens are included as word boundaries, which determines how words are
counted in situations such as {\em PPP\/} in the next line of
Figure~\ref{fig:codings}.  We also count slashes as boundaries, so that the
three acronyms {\em MIT\/}, {\em LCS\/} and {\em TR\/} can be extracted from
{\em Laboratory for Computer Science, Massachusetts Institute of Technology,
Technical report}.  We also delete numbers, which affects the counting of words
in the final {\em UW\/} (and the {\em FG\/}) example in the Figure.

\begin{figure}
\footnotesize
\begin{center}
\begin{tabular}{l l l}
\hline
BC	& Both the Bandwidth Contraction (BC) algorithm
	& \verb.-  2 <1> <1,1>.						\\
OTC	& OTC represents one time costs that are not spread
	& \verb.+  2 <1,1> <1,1,1>.					\\ 
FG	& For fast Givens transformations Equation 2
	& \verb.-  7 <1> <1,1>.						\\
	& \hspace{0.2cm} becomes Total flops FG = 23 ( Ttotal ) +
	&								\\
CHARME	& CHARME '93: IFIP WG10.2 Advanced Research
	& \verb.+  8 <1,4> <1,3,2>.					\\
	&  \hspace{0.2cm} Working Conference on Correct Hardware Design and
	& 								\\
	& \hspace{0.2cm} Verification Methods
	&								\\
CDAG	& the cluster dependency DAG (CDAG)
	& \verb.-  3 <2> <1,3>.						\\
COMPCON & In Procedures of the IEEE Computer Society
	& \verb.-  4 <3> <4,3>.						\\
	& \hspace{0.2cm} International Conference (COMPCON)
	&								\\
OOPSLA	& In N. Meyrowitz, editor, Object-Oriented
	& \verb.-  7 <1,1,1,1,2>.					\\
	& \hspace{0.2cm} Programming Systems, Languages and Applications
	& \hspace{0.2cm} \verb.   <1,1,1,1,1,1>.			\\  
	& \hspace{0.2cm} (OOPSLA '86)			&		\\
PPP	& the Point-to-Point Protocol (PPP) [Sim93, McG92] 
	& \verb.-  4 <2,1> <1,1,1>.					\\ 
LCS	& Ph.D. thesis, Laboratory for Computer Science,
	& \verb.- 11 <2,1> <1,1,1>.					\\ 
MIT	& \hspace{0.2cm} Massachusetts Institute of Technology, Technical
	& \verb.-  6 <1,2> <1,1,1>.					\\ 
TR	& \hspace{0.2cm} report MIT/LCS/TR-354 &  \verb.-  4 <1> <1,1>.	\\
TR	& Also available as Technical Report TR94-1468
	& \verb.-  2 <1> <1,1>.						\\ 
UW	& Technical Report UW-CSE-94-11-08, University of
	&  \verb.+  2 <2> <1,1>.					\\ 
	& \hspace{0.2cm} Washington &					\\
\hline
\end{tabular}
\end{center}
\caption{Acronyms, their contexts, and their codings with respect to the context}
\label{fig:codings}
\end{figure}
Our coding scheme does not permit certain acronyms to be encoded.  Some
examples from the data we used are shown in Figure~\ref{fig:unencodable}.  We
insist that words appear in the correct order, ruling out the encoding of {\em
ISO\/} in the first example.  The English expansion of foreign language
acronyms like {\em NWO\/} often does not include the letters of the acronym.
Plural forms cause havoc, whether capitalized ({\em PITS\/}) or not ({\em
CVEs\/}).  Occasionally an acronym comes in the middle of its definition, as
with the {\em SIS\/} example.  Sometimes letters are plucked out of the middle
of words, as in {\em NEBW\/} and {\em JPTN}.  Not only does the acronym {\em
B8ZS\/} include a digit, but it requires the domain knowledge that the
character string ``eight'' corresponds to the character ``8''.  All these
examples will count as failures of our acronym extraction procedure, although
it should be noted that only a few of them satisfy Webster's definition of an
acronym in the strict sense.

\begin{figure}
\footnotesize
\begin{center}
\begin{tabular}{l l}
\hline
ISO & International Organisation for Standardisation document ISO/IEC JTC1/SC29
	\\
NWO & the Netherlands Organization for Scientific Research (NWO) \\
PIT & two considers Populated Information Terrains (PITS) \\
CVE & so called Collaborative Virtual Environments (CVEs) \\
SIS & and Shared Interface (SIS) Services, prototyped in the work of strand 4 \\
NETBW & by the network bandwidth (NETBW ) \\
JPTN & A Jumping Petri Net ([18], [12]), JPTN for short \\
B8ZS & Bipolar with eight zero substitution coding (B8ZS)  \\
\hline
\end{tabular}
\end{center}
\caption{Acronyms in contexts that cannot be encoded}
\label{fig:unencodable}
\end{figure}

\subsection{Compressing the acronym representation}

In order to compress the candidates, we code the acronym encodings exemplified
in Figure~\ref{fig:codings}.  There are four components: the direction, the
first-word offset, the subsequent-word offsets, and the number of characters
taken from each word.  Different models are used for each, simple zero-order
models in each case, formed from the training data.  New acronyms are encoded
according to these models using arithmetic coding; a standard escape mechanism
is used to encode novel numbers appear.

After compressing the acronym candidates with respect to their context, all
legal encodings for each acronym are compared and the one that compresses best
is selected.  (In the event of a tie, both are selected.)  We then compress the
acronym using the text model, taking the preceding context into account but
only calculating the number of bits required to represent the characters in the
acronym.

The two compression figures are compared, and the candidate is declared to be
an acronym if
\[ \frac{\rm bits_{~acronym~model}}{\rm bits_{~PPM~model}} \leq t \] 
for some predetermined threshold $t$.  Note that subtracting the number of bits
is more easily justified than using the ratio between them, because the
difference in bits corresponds to a likelihood ratio.  In fact, however, far
better results were obtained using the ratio method.  While we do not fully
understand the reasons for this, it is probably connected with the curious fact
that longer acronyms tend to compress into fewer bits using a standard text
model than shorter ones.  While short acronyms like {\em BC} or {\em PPP} are
often pronounced as letters, long ones like {\em CHARME\/}, {\em COMPCOM} and
{\em OOPSLA} tend to be pronounced as words.  This affects the choice of letter
sequences that are used: longer acronyms tend to more closely resemble
``natural'' words.

\section{Experimental results}

To test these ideas, we conducted an experiment on a sizable sample of
technical reports.

\subsection{Data}
\label{sec:data}

We used as training and test data 150 reports from the {\em Computer Science
Technical Reports\/} collection of the New Zealand Digital
Library.\footnote{{\em www.nzdl.org}} These have been extracted automatically
from PostScript files, and contain a certain amount of noise that can be
attributable to errors in that process.  The total size of this corpus is 9.3
Mb, or 1.4 million words.  Two-thirds of the documents are used for training
and the remainder for testing.

Approximately 1080 acronym definitions have been identified manually in the
training and test documents.  (In fact, a semi-automated process was used for
some, but they have all been checked manually).  The acronyms range from two to
seven letters in length (single-character abbreviations are not counted as
acronyms).  Approximately 600 are two-letter acronyms, and there is only a
sprinkling of six- and seven-letter acronyms.

Of the 440,000 words that appear in the test documents, 10,200 are upper-case
words of two characters or more---and thus candidates for acronym definitions.
Of these, only 10.6\% are actually acronym definitions.

\subsection{Evaluation}

In order to evaluate acronym identification schemes, we face a standard
tradeoff between liberal algorithms that increase the chance that a particular
acronym definition is spotted but also increase the number of ``false
positives,'' that is, other segments of text that are erroneously flagged as
acronym definitions; and conservative algorithms that reduce the number of
false positives but also increase the number of ``false negatives,'' that is,
acronym definitions that are not identified as such by the system.  This
tradeoff is familiar in information retrieval, where a search engine must
decide how long a list of articles to present to the user, balancing the
disadvantage of too many false positives (irrelevant documents that are
displayed), if the list is too long, against that of too many false negatives
(relevant documents that are not displayed), if it is too short.

Following standard usage in information retrieval, we quantify this tradeoff in
terms of ``recall'' and ``precision'':
\[ {\rm recall} =
     { { {\rm number~of~test~articles~correctly~assigned~to~category~} C}
       \over {\rm total~number~of~test~articles~that~have~category~} C }  \]
\[ {\rm precision} =
     { { {\rm number~of~test~articles~correctly~assigned~to~category~} C}
       \over { {\rm total~number~of~test~articles~to~which~category~} C
               {\rm ~is~assigned} } } .  \]

\subsection{Results}
\label{sec:results}

Acronyms with just two letters are significantly more difficult to extract than
longer acronyms, both because there is a greater probability that a random
sequence of words will appear by coincidence to be an acronym definition, and
because there is more opportunity to pluck the wrong sequence of words out of a
legitimate acronym definition.  Consequently we look separately at results for
acronyms for two or more letters and for ones of three or more letters.

\begin{figure}
\begin{center}
\begin{tabular} {c c}
\epsfig{file=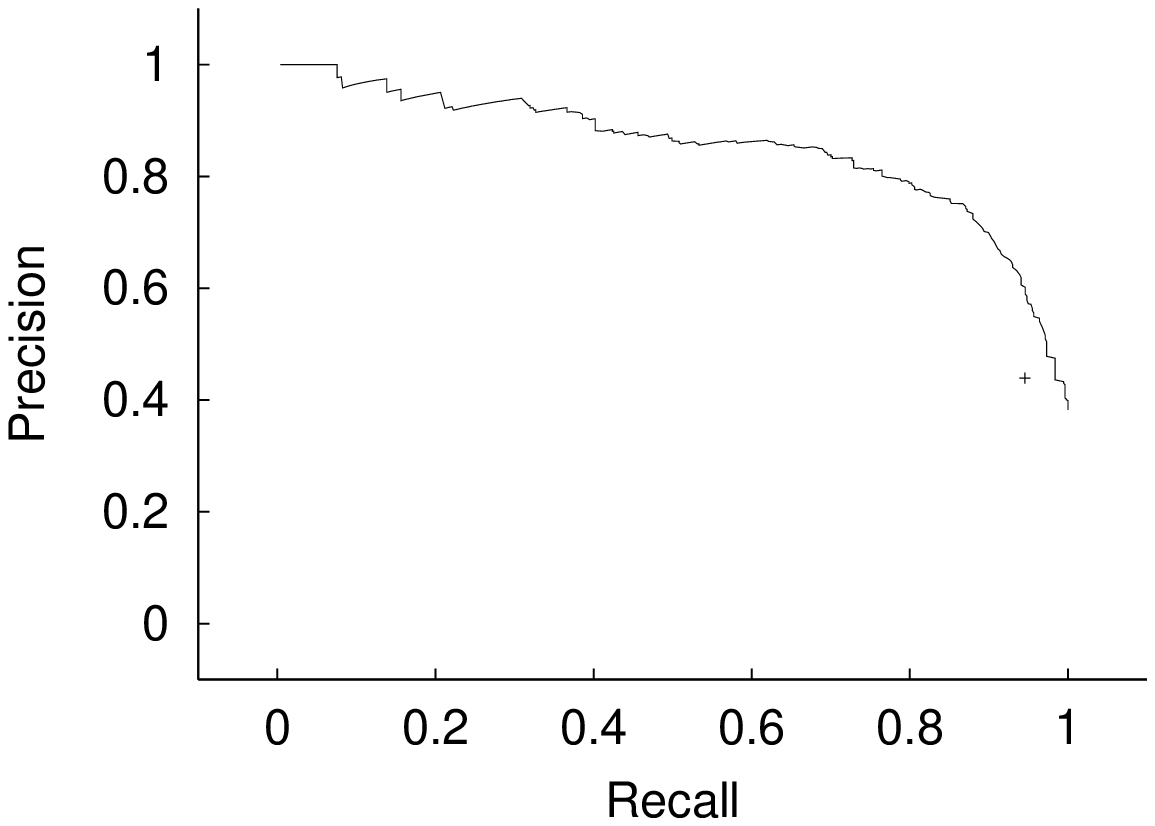,width=3in,angle=0}
 & \epsfig{file=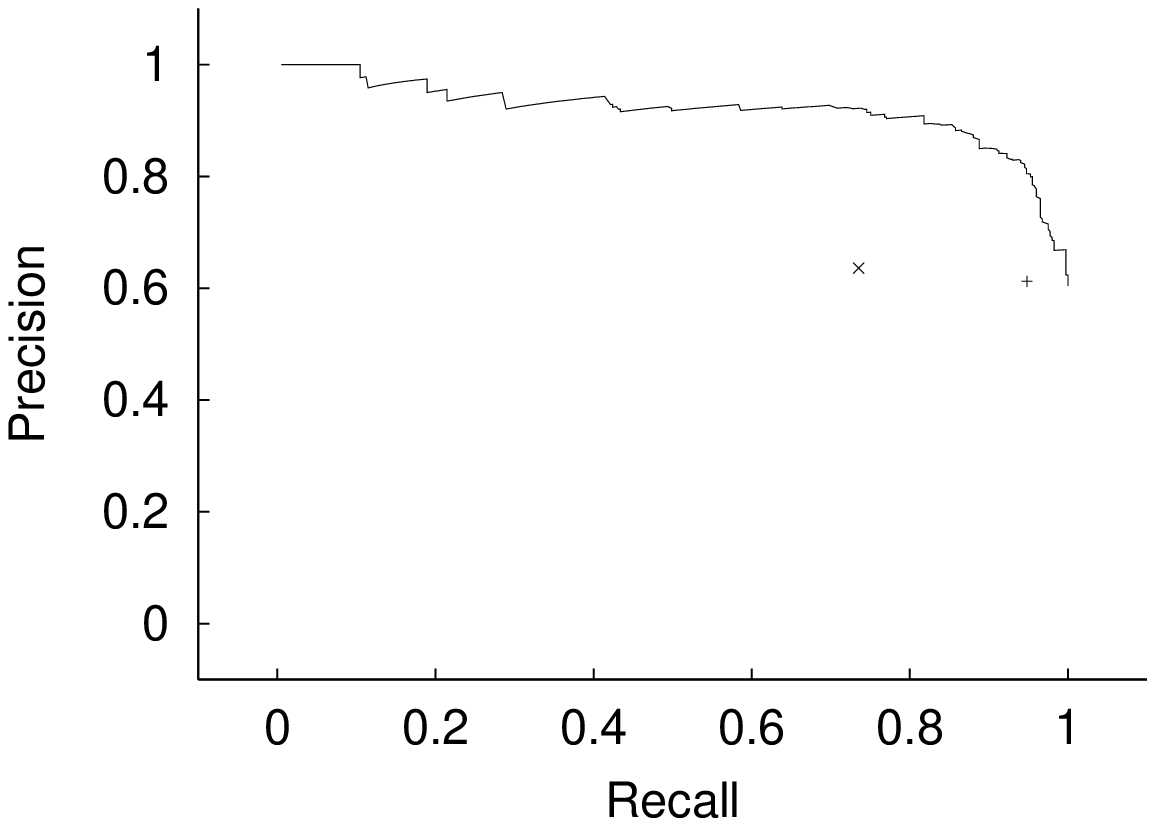,width=3in,angle=0} \\
(a) & (b)
\end{tabular}
\end{center}
\caption{Recall--precision curves for compression-based acronym extraction:
         (a) acronyms of two or more letters;
	 (b) acronyms of three or more letters.}
\label{fig:recallprecision}
\end{figure}

Figure~\ref{fig:recallprecision} shows recall-precision curves for
compression-based acronym detection in the two cases.  The curves are generated
by varying the threshold $t$ that governs the acceptance of an acronym.  The
lefthand end corresponds to the small value of $t = 0.1$.  Here a candidate
acronym is accepted if the acronym model compresses it to less than 10\% of the
amount achieved by PPM.  Very effective compression is a strong indicator of
the presence of an acronym, leading to high precision---but, unfortunately,
very low recall.

As the threshold increases from $t = 0.1$, recall steadily improves.  There is
a long flat section as $t$ increases to about $0.2$, with a precision that
remains constant at 85\%--90\% for acronyms of three or more letters at recall
values up to almost 80\% (Figure~\ref{fig:recallprecision}b).  When $t$
increases beyond about $0.2$ (which occurs in Figure~\ref{fig:recallprecision}b
at a recall level of about 80\%), recall continues to rise but precision begins
to tail off, due to the rising probability of naturally occurring sequences
that look like acronym definitions but are not.  For a sufficiently large value
of $t$, all candidate acronyms that can be made up of initial letters taken in
order from words in the surrounding context are detected.  This corresponds to
a recall of 1, and a precision of about 60\% for three--or-more-letter acronyms
and 30\% for two-or-more-letter acronyms.  Finally, some of the noise in
Figure~\ref{fig:recallprecision}a and \ref{fig:recallprecision}b is caused by
multiple occurrences of certain acronyms in some documents---for example, one
document defines {\em MIME (Multipurpose Internet Mail Extensions)} in its page
header, giving a total of 67 occurrences.

The points marked $+$ close to the line in Figures~\ref{fig:recallprecision}a
and \ref{fig:recallprecision}b mark the recall--precision point for the TLA
heuristic extraction algorithm.  The compression-based method is a clear
improvement on TLA in both cases.  The lone point marked $\times$ shown in
Figure~\ref{fig:recallprecision}b is the result of the simple Perl heuristic,
which performs much more poorly.

The only results reported for AFP are for a corpus with radically different
characteristics to ours.  Taghva and Gilbreth (1995) used government
environmental studies with an average of 27 acronyms per document, whereas we
used computer science technical reports with an average of 8.5 acronyms per
document.  Thus the results are not comparable.

\section{Conclusions}

Compression-based text mining seems to provide a viable basis for extracting
acronyms and their definitions from plain text.  It gives an advantage over
other methods in that it reduces the need to come up with heuristics for
deciding when to accept a candidate acronym---although some prior choices must
be made when designing the method for coding acronyms with respect to their
definitions.  It also allows the operator to choose any point on the
recall-precision curve.

This work represents an initial foray into compression-based acronym
extraction.  There are three obvious areas for further improvement.  First,
case information in acronym definitions, currently ignored, could usefully be
exploited.  Authors regularly capitalize initial letters that occur in acronym
definitions, as in almost all of the examples in Figures~\ref{fig:codings} and
\ref{fig:unencodable}, or emphasize them by other means, like italics, as in
Webster's definition with which we began.  In many applications, font
information such as italics will not be available, but capitalization will be.
This information can be accommodated by incorporating a capitalization flag
into the encoding scheme.  Capitalized letters in the middle of words are also
a good indication that they participate in an acronym definition, and should
also be incorporated into the encoding.

Second, the appearance of parentheses---or occasionally quotation
marks---around either the acronym or its definition (whichever appears last) is
another indicator that is presently being ignored.  It appears in around
two-thirds of the examples in Figures~\ref{fig:codings} and
\ref{fig:unencodable}---in every case around the acronym rather than its
definition.  This could allow us to confirm the presence of an acronym and
permit lower-case acronyms, albeit rare, to be spotted.  In those occasions
where parentheses appear around the definition they could be encoded as
begin-definition and end-definition items.

Finally, the compression-based method as currently implemented suffers because
it does not accommodate approximate matching of the acronym with its
definition.  This could be incorporated using standard zero-frequency
mechanisms, although that would increase the search space and slow down acronym
detection substantially.

\subsection*{References}
{\small\frenchspacing
\begin{list}{}{\setlength{\itemsep}{0mm}%
                \setlength{\parsep}{0mm}}

\item[Bell, T.C., Cleary, J.G. and Witten, I.H. (1990)] {\em Text
compression\/}.  Prentice Hall, Englewood Cliffs, NJ.

\item[Cormen, T.H., Leiserson, C.E.~and Rivest, R.L. (1990)] {\em Introduction
to algorithms\/}.  MIT Press, Cambridge, MA.

\item[Howard, P.G. (1993)]
``The design and analysis of efficient lossless data compression systems.''
Ph.D.~thesis, Brown University, Providence, Rhode Island.

\item[Taghva, K. and Gilbreth, J. (1995)]
{\em Recognizing acronyms and their definitions.}  Technical Report
Taghva95-03, ISRI ISRI; November.

\item[Witten, I.H., Bray, Z., Mahoui, M. and Teahan, W.J. (1999)] ``Text
mining: a new frontier for lossless compression.''  {\em Proc Data Compression
Conference\/}, pp.~198--207.  IEEE Press, Los Alamitos, CA.

\item[Witten, I.H. and Frank, E. (2000)] {\em Data mining: Practical machine
learning tools and techniques with Java implementations\/}.  Morgan Kaufmann,
San Francisco, CA.

\item[Yeates, S. (1999)]
``Automatic extraction of acronyms from text,'' {\em Proc New Zealand Computer
Science Research Students' Conference\/}, pp.~117--124.  University of Waikato,
Hamilton, New Zealand.

\end{list} }

\end{document}